# A PROTOTYPE DATA ACQUISITION AND PROCESSING SYSTEM FOR SCHUMANN RESONANCE MEASUREMENTS

Giorgos Tatsis[a], Constantinos Votis[a], Vasilis Christofilakis[a,•], Panos Kostarakis[a], Vasilis Tritakis[b], Christos Repapis[b]

[a]Physics Department, Electronics-Telecommunications and Applications Laboratory, University of Ioannina, Ioannina, Greece

[b]Mariolopoulos-Kanaginis Foundation for the Environmental Sciences

**Abstract-** In this paper, a cost-effective prototype data acquisition system specifically designed for Schumann resonance measurements and an adequate signal processing method are described in detail. The implemented system captures the magnetic component of the Schumann resonance signal, using a magnetic antenna, at much higher sampling rates than the Nyquist rate for efficient signal improvement. In order to obtain the characteristics of the individual resonances of the SR spectrum a new and efficient software was developed. The processing techniques used in this software are analyzed thoroughly in the following. Evaluation of system's performance and operation is realized using preliminary measurements taken in the region of Northwest Greece.

.

## 1. Introduction

The space between the earth and the terrestrial ionosphere forms a large electromagnetic waveguide, considering that the earth surface is a conducting sphere as the ionosphere boundary. It is experimentally proven that the earth-ionosphere waveguide permits the generation of electromagnetic standing waves in the range of extremely low frequencies (ELF), where the wavelength is related to the dimensions of the earth-ionosphere cavity (Balser & Wagner, 1960; Balser & Wagner, 1962).

[•] Corresponding Author: e-mail: vachrist@uoi.gr, vasilis.christofilakis@gmail.com ; Tel.: +30 26510 08542 ; fax: +30 26510 08674

Therefore, specific spectral resonant frequencies are observed. The existence of these resonances was first predicted by Schumann (1952) on a theoretical basis and since then we refer to this phenomenon by the term Schumann resonance (SR) (Schumann, 1952). The frequencies of the lower eigenmodes are approximately 7.8, 14, 21, 27, 33 Hz. It is found that the main source of these electromagnetic waves is the lightning effect, occurring on the earth. Consequently, Schumann resonance measurements could be used to monitor the global lightning activity during day and seasons (Heckman, Williams, & Boldi, 1998; Nickolaenko & Rabinowicz, 1995; Nickolaenko, 1997). Several natural phenomena such as the climate variations, ionosphere disturbances, solar radiation, have an impact in the spectrum of Schumann resonances. Schumann resonance signals could be the global environmental signal absorbed by the human body (Palmer, Rycroft, & Cermack, 2006). Furthermore, the characteristics of the SR may be important in aerospace, marine applications (Tulunay et al., 2008). A number of corresponding researches have been conducted in order to investigate several types of natural events through spectral traces. To report some of them, global temperature variations, water vaporization, space weather and earthquake precursors could be monitored by SR spectrum measurements (Sekiguchi, Hayakawa, Nickolaenko, & Hobara, 2006; Roldugin, Maltsev, Petrova, & Vasiljev, 2001; Hayakawa, Ohta, Nickolaenko, & Ando, 2005; Hayakawa et al., 2010). The significance of Schumann resonance spectrum has attracted the scientific interest of researchers around the world. Until today, several scientific groups have taken measurements of the SR spectrum and stations have been installed in order to record the spectral variations in daily basis (Sierra et al., 2014; Toledo-Redondo et al., 2010). Besides the important role of SR spectrum measurements for the scientific community, there are additional motivations for this research:

- Firstly was the comparison of simulation results with preliminary measurements taken occasionally, from August 2013 to December 2014, on the field comparing in this way the functionality of the SR measuring equipment.
- Secondly was the implementation of a prototype portable low cost, scalable and effective data acquisition system (DAS) capable to capture signals in the extremely low frequency (ELF) and the lower portion of the very low frequency (VLF) band. Exploiting the benefits of adjustable sampling rate, it is feasible to oversample the lowest part of ELF between 0 to 30Hz where SR frequencies lie around.
- Last but not least was to carry out measurements in a location that was never measured, like Greece.

The final objective of current research is the implementation of a fully autonomous SR recording station, as a part of SR network, able to capture continuously and monitor special characteristics of the spectrum, along with its variations over time. The SR detection system consists of three main parts that are the antenna the amplification and filtering and the DAS. The latter includes a prototype data logging device and a personal computer. The rest of this paper is organized into the following sections; Section 2 presents related works; Section 3 gives a brief description on the analog part of our system including the antenna used and the amplifier-filters chain. In section 4 the DAS is described analytically. Section 5 is dedicated to the results and the signal processing carried out by the software. Conclusions and future work are drawn in section 6.

## 2. Related Works

Capturing of naturally occurring electromagnetic events such as Schumann resonances in the portion of ELF (0-3 KHz) and VLF (3-30 KHz) has not been widely covered (Hanna, Motai, Varhue & Titcomb, 2009). There is still a lack of reliable, economical technologies for monitoring the RF of the lithosphere-ionosphere coupling processes (Yi & Liu, 2011). Focusing mainly on the data acquisition component of SR monitoring systems previously works are presented. At Modra observatory, the output signal is digitized by 16-bit ADC at a sampling rate of 200Hz. Authors have made several measurements from sampling 180 up to 230Hz to check possible aliasing effects (Ondrášková, Ševčík & Kostecký, 2009) A commercial DAS with an adjustable sampling rate up to 4 KHz was used as a part of a new observation system to measure magnetic fields in the extremely low frequency band at Moshiri, Hokkaido, Japan. The commercial data acquisition board is placed in a PCI slot of a PC 220m away from the antenna (Ando et al.2005). At a vast bare land near Kolkata two research teams from India are capturing the ELF signals at a rate 40000 times per second using a 12-bit A/D. Information is lost in the sampling process only due to relatively small A/D's bit resolution (De et al., 2010). Schumann resonance frequency variations observed in the Himalayan region, India, at elevations between 1228–2747 electromagnetic field components, in the form of time series, were recorded at 64 Hz sampling frequency at a site located away from the cultural noise (Chand, Israil & Rai, 2009). One research team from the University of Electro-Communications, Japan and Usikov's Institute for Radiophysics and Electronics, Ukraine are digitized signals inside the ELF band at 350Hz using 12-bit A/D (Shvets, Hayakawa, Sekiguchi & Ando, 2009).

According to previous works, there is an extra need for portable, autonomous, low-cost, data acquisition systems capable of capturing signals in the ELF and VLF band. The present prototype data acquisition and processing system presented in this work as a part of SR measurement system. Data logger is completely autonomous for a time period of one month concerning either storage or battery power capacities. The adjustable sampling rate from 512Hz up to 16384Hz enables oversampling of the 30Hz Schumman bandwidth even with a 273 times ratio. Oversampling in combination with 16-bit resolution is giving efficient signal detection. Furthermore oversampling eliminates the necessity of anti-aliasing filter at the analog front end after the antenna. Additionally the upper limit of sampling at 16384Hz gives the ability to capture signals inside the entire ELF band and at the lower band of VLF (up to 8KHz). Operation of the system using preliminary measurements is presented in details in the next sections.

**3. Antenna and signal conditioning**

The architecture of the Schumann resonances detection system we used in experimental measurements is based on a magnetic field antenna and the corresponding signal conditioning system. The first implementation is an inductor coil with a ferromagnetic core material that exhibits relative magnetic permeability value of the order of $10^5$. That core material is a mumetal which meets ASTM A753 Alloy 4 specifications. The proposed inductor coil has 40,000 number of turns on a mumetal core rod (length 300 mm & diameter 25 mm). The induced voltage amplitude at the inductor coil terminal exhibits extra low values due to the extra low magnetic field amplitudes (some tenths of pT) that provide Schumann resonances. In order to achieve signal conditioning on those low-level signal amplitudes ensuring efficient noise performance and less signal distortion and degradation we designed and implemented a number of amplifying and filtering circuitries that are embodied in cascading topology. That proposed signal conditioning system provides remarkable performance at a wide frequency range. Amplifying, filtering and noise aspects on electronic circuitries have in detail studied and investigated in order to achieve system performance optimization. The overall gain of the amplification is 118dBV. The cutoff frequency (at -3dB) is 27Hz, whereas the rolloff is about 60dB/octave. Moreover, we managed to eliminate the signal level of the 50 Hz parasitic single tone from household supply system at the signal conditioning system output. This is achived by using a hardware notch filter as part of the rest signal conditioning system. Additional information about the antenna and the signal conditioning are presented in our previous work (Votis et al., 2015). Fig.1 depicts a photo of the Schumann resonances detection system.

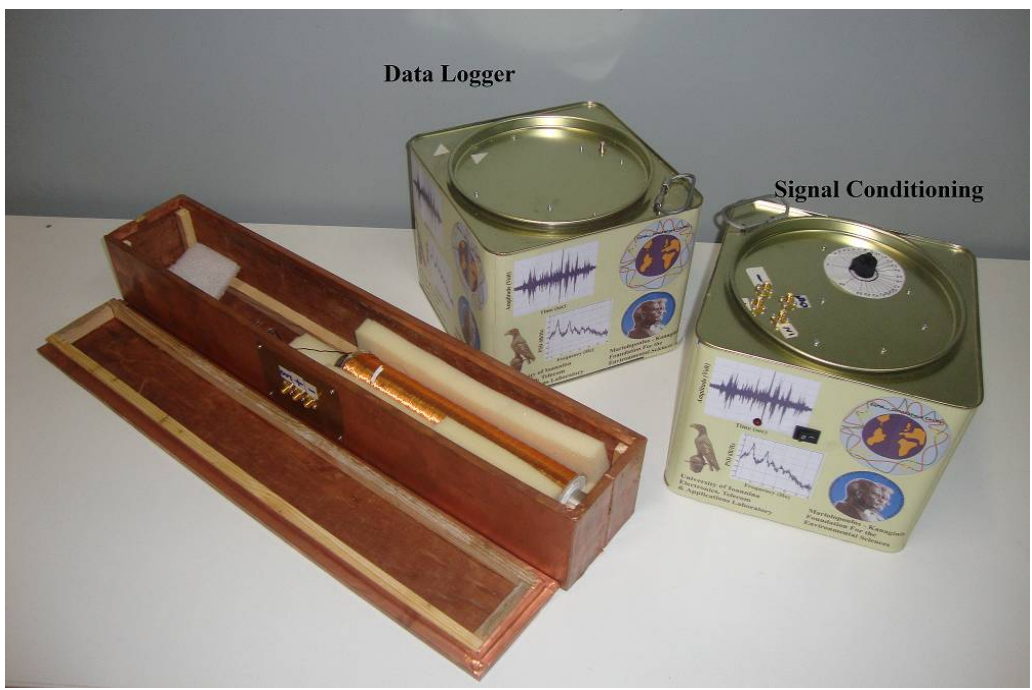

**Fig 1. Photograph of the proposed Schumann resonance detection system**

## 4. Data acquisition system overview

### 4.1 Architecture

In order to analyze the analog signal captured by the magnetic sensor, this signal is recorded in digital form and then the new off-line software processes it. The digitization and storage are carried out by a data logging device that has been designed and constructed for this purpose. The logger consists of three main parts, an analog to digital converter (ADC), a central processing unit; that is a microcontroller (MCU) and a storage device. A simplified block diagram is shown in Fig. 2. The ADC converts the received analog signal to digital and via the MCU it is stored in a non-volatile memory device. The storage device used is a Secure Digital High Capacity flash memory (SDHC) capable of saving binary data of several gigabytes. Our data logger is also equipped with a communication bus. The communication bus allows interconnection with a PC and is used for initialization of the data logger before its stand-alone use. The same communication bus also permits the use of our data logger as a spectrum analyzer when connected to a PC. This feature is very useful for calibrating the analog chain, measuring the gain, the precise response of the filters e.t.c.

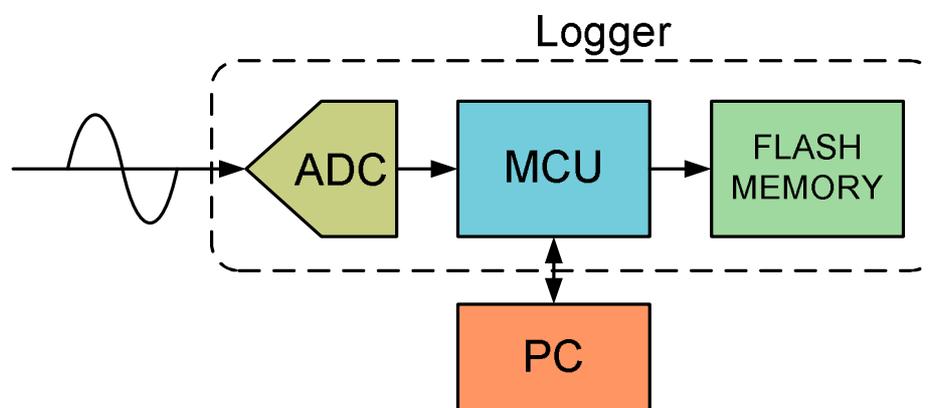

**Fig 2. Data acquisition block diagram**

**4.2 Data logger design and specifications**

Prior to the design of this device, special care was taken in order to specify the appropriate characteristics of the logging procedure. The most important parameters, which must be determined, are the resolution of the A/D converter and the sampling rate. A/D sampling rate as well as bit resolution was evaluated through several simulations. The simulation procedure is as follows; first an analog signal is constructed containing five frequencies near the first five expected Schumann resonances that are approximately, 7.8Hz, 14Hz, 21Hz, 27Hz, 33Hz. The signal has duration of 10 minutes. This duration is considered to be long enough in order to efficiently estimate the power spectral density of SR signal in the presence of strong noise and at the same time short enough in order to consider that the resonances are constant during this time (Nickolaenko & Hayakawa, n.d.). After that, a white Gaussian thermal noise is added and then the whole signal (analog signal + noise) is quantized according to the A/D bit resolution. Finally, the signal's corresponding magnitude spectrum is captured by using the fast discrete Fourier transformation (FFT). In order to reduce the noise, the FFT is not performed on the entire signal duration (10min) but in segments of 30 seconds. The individual spectrums, of these segments, are averaged similarly to the Bartlett's method (Proakis & Manolakis, 1996; Babtlett, 1948). It is well known that during the digitization of an analog signal, a quantization error is introduced. The quantization error is derived from the difference between the real value analog signal and the corresponding integer approximation represented by the available A/D output bits (b). If the analog signal amplitude does not exceed the full scale range of the A/D and the quantizer error is uniformly random variable then the quantization noise power is given by (Bennett, 1948):

$$N_q = \frac{\Delta^2}{12R},\qquad(1)$$

Where $\Delta = \frac{V_{pp}}{2^b}$ stands for the A/D bit size, $V_{pp}$ is the peak to peak amplitude at the A/D input and R is the input resistance of the A/D. For the case of simplicity we consider that R=1Ω. From the Eq.1 it is clear that the quantization noise power depends only on bit resolution of the A/D. Specifically as the number of bits increases, the quantization error decreases. (Gray & Neuhoff, 1998).

Fig. 3. shows the simulated Schumann spectrum, obtained with the aforementioned procedure, without the addition of thermal noise. The depicted noise is a result of quantization error only. Simulations show that as the number of the bits increases the results are better as expected and a higher resolution A/D is a great benefit. However the difference between 16bits and 24bits is quite small and the improvement is considered negligible. Therefore we choose the 16bits for the resolution of our A/D as the best compromised solution between noise-performance and efficiency.

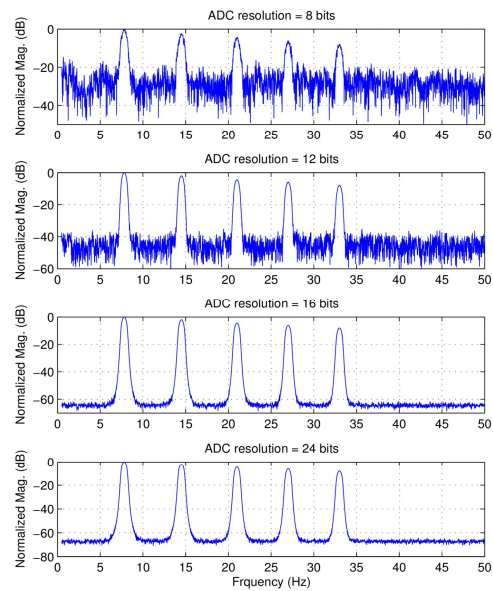

**Fig. 3. Simulated Schumann spectrum against different A/D bit resolution without additive noise**

At this paragraph, the benefits of oversampling are underlined (Candy & Temes, 1992). For an uniform A/D with sampling at the Nyquist rate the spectral power density of a bandlimited signal with a maximum frequency component $f_B = \frac{f_s}{2}$ is shown in Fig. 4(a). The following equation gives the spectral power density at the Nyquist rate (Hauser, 1991), (Widrow & Kollar, 2008, pp. 529-562).

$$PSD_{Nyquist}(f) = \frac{\Delta^2}{12 f_s} = \frac{\Delta^2}{12} \frac{1}{2 f_B} \qquad (2)$$

Suppose that a sampling rate $f_s'$ much higher than the Nyquist rate is used. The

$$PSD_{oversampling}(f) = \frac{\Delta^2}{12 f_s'} \qquad (3)$$

along with PSD$_{Nyquist}$ is illustrated in Fig. 4(b)

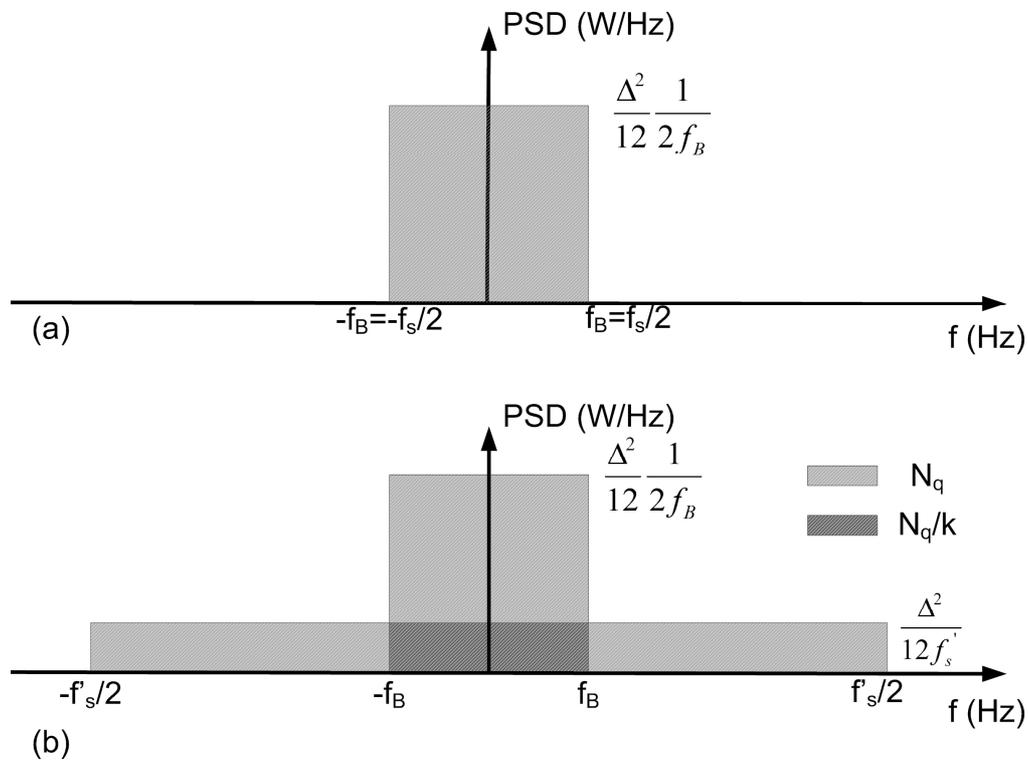

**Fig. 4. Graphical representation of the oversampling benefits over noise**

For the case of oversampling, quantization noise inside our bandwidth (-f$_B$ to f$_B$) is given by:

$$N = \int_{-f_B}^{f_B} \frac{\Delta^2}{12 f'_s} df = \frac{\Delta^2}{12} \frac{2 f_B}{f'_s} = N_q \frac{2 f_B}{f'_s} = \frac{N_q}{k} \quad (4)$$

Sampling at rates higher than the Nyquist rate reduces the amount of quantization noise inside our bandwidth (-f$_B$ to f$_B$) as highlighted on Fig.4(b). The amount of reduction is defined by the parameter $k = \frac{f'_s}{2 f_B}$, commonly known as oversampling ratio. Assuming a full-scale sine wave input, then signal's power equals:

$$P = \left( \frac{2^b \Delta}{2\sqrt{2}} \right)^2 \quad (5)$$

The Signal to Noise Ratio (SNR) is given by:

$$SNR = 10 \log\left(\frac{P}{N}\right) = 10 \log\left(\frac{3}{2} 2^{2b} k\right) = 10 \log\left(\frac{3}{2}\right) + 10 \log(2^{2b}) + 10 \log(k) \Rightarrow$$
$$SNR = 6.02 b + 1.76 + 10 \log(k) \quad (6)$$

The $10 \log(k)$ term, in Eq.6, indicates that it is preferable to use sampling rates as much higher than the Nyquist rate as possible (oversampling) because the SNR regarding quantization is increased. An increase of 2 bits in the bit resolution implies an increase of the SNR of approximately 12 dB. The same improvement in the SNR is obtained when the oversampling ratio k=16 (Christofilakis, Alexandridis, Kostarakis, & Dangakis 2002; Christofilakis, Alexandridis, Dangakis, & Kostarakis, 2003).

The logger's sampling rate is selectable by the user and it takes several fixed values from 512Hz up to 16384Hz.

Fig 5. shows a photograph of the constructed logger and Table 1 shows the technical specifications of this device. Programming of our data logger, as well as setting of the parameters such as record length, sampling rate e.t.c., is performed from a PC via USB port that allows communication with our graphical user interface (GUI). This initialization is performed once before running and then the data logger is a standalone device that does not need to be connected to a PC. During power-on, the data logger restores the last saved settings from the embedded EEPROM (Electrically Erasable Programmable Read-Only Memory). The communication between the device and the PC can also be used for real-time testing measurements on the received analog signal (time-series) offering an oscilloscope-like functionality. As mentioned above, the data sampled by the logger are stored on a flash memory card. The storage is carried out organized in separate files. As mentioned before the user is allowed to control the time-length of these files via the GUI (Fig. 6). For example, one could set this length to 10 minutes. In other words, the device is sampling data continuously for 10 minutes in one file and then start saving the following incoming data into a new file for the next 10 minutes. In addition, a real time clock module (RTC) is added to record the date/time of the captured data stored in the files. This timestamp allows the user to monitor the variation of the measured spectrum during the time in user defined intervals.

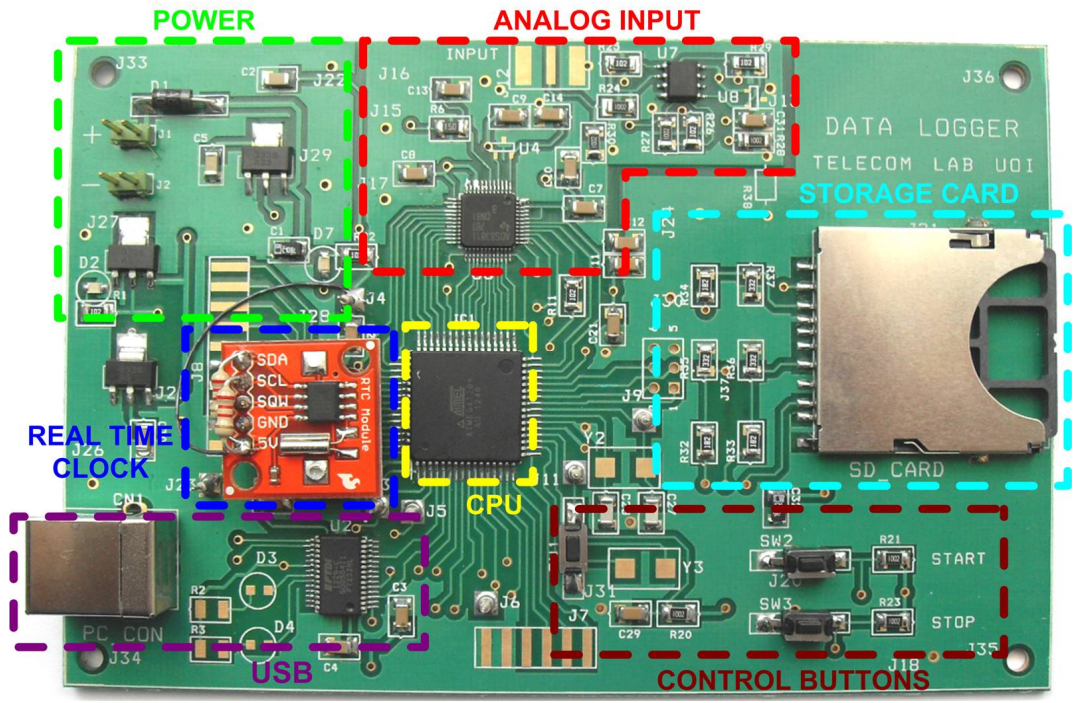

**Fig. 5. Data logger photograph**

| Table. 1. Logger Technical Details | |
|---|---|
| A/D Converter | ADS8381 (max. rate 580 KS/sec) |
| Microcontroller | ATMEGA128 |
| Real Time Clock | DS1307 |
| Storage slot | SDHC card, FAT32 |
| PC Connection | USB 1.1 spec. |
| Power | 12V, ~80mA |
| Autonomy — Power | ~35 days with a 70Ah battery (can be extended with use of solar panels) |
| Autonomy — Data | ~40 days, with 64GB card (for $f_s$=9600 Hz, ~1.545GB per day) |

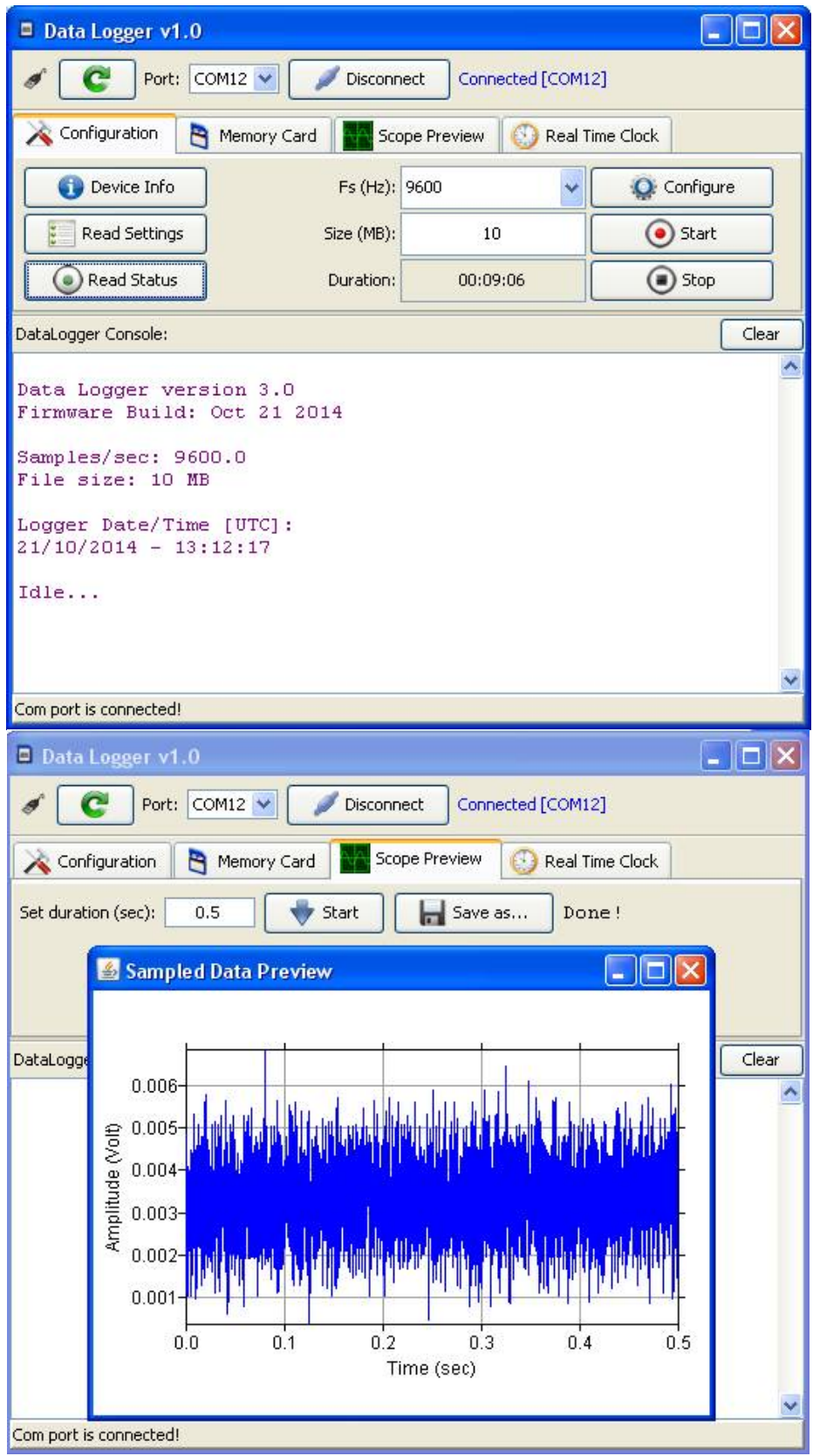

**Fig. 6. Data acquisition graphical user interface (GUI)**

## 5. Signal processing and results

All experiments and preliminary measurements took place from August 2013 to December 2014 in several spots in the region of Epirus (NW Greece). The main reason for choosing the spot with Latitude 39°30'38.3"N and Longitude 21°00'31.3"E was the far distance from electromagnetic pollution sources and specifically from the public electrical network (Fig. 7). The raw digital data recorded by the logger at the mentioned spot represent the analog signal captured by the magnetic sensor after adequate amplification and filtering. The final stage of our measuring system is the processing of this signal in order to obtain the corresponding Schumann resonances. The signal processing is carried out entirely via software running on a personal computer (PC). The goal is twofold. First, we need to obtain an estimate of the spectrum of the signal. Second, we focus on the particular characteristics of the expected resonances in order to estimate the parameters of these resonances. For the first part, we use the Welch method to estimate the power spectral density (PSD) of the signal (Welch, 1967). Fig.8. shows an example of actual measurement that took place outdoors, during measurements of SR with our measuring system. On the top, one can see the sampled signal received from the sensor after proper amplification and filtering. In fact, this is the analog representation of the digitized signal. The vertical axis is appropriately scaled in order to convert the 16bit wide digital samples from the A/D into the corresponding analog signal's Volts. The logger is sampling the analog signal at a rate of 9600 samples/sec for 10 minutes. The A/D resolution is 16 bits. On the bottom side of Fig. 8. we see the PSD as calculated using the Welch method. It is well known that this method is the most appropriate to be used for very noisy signals. For the spectral estimation, the initial time-signal is divided into segments of 30 seconds duration, which are overlapped each other with 50% coverage percentage. The segments are windowed before the Discrete Fourier Transformation (DFT), using a Hamming windowing function. The frequency resolution produced is 0.033Hz, according to the segments duration. Three peaks near 8Hz, 14Hz and 20Hz are visible. The bandpass-like shape is due to the filtering used. A well-known peak, strictly 50Hz, appears in the spectrum that is induced by electricity lines transfer that lie at a distance of 2.5 Km away from the sensor (Fig. 7). Several preliminary measurements have taken place at the same location of previous photograph on different dates as Fig. 9 shows. By observing the depicted spectrums it is clear that at least the three first schumann resonances are visible. There are some variations of the magnitudes and the frequencies of the resonancies over time. Also the power lines frequency of 50Hz is visible. The 50Hz peak is a very strong interference, but it is actually attenuated by the filters.

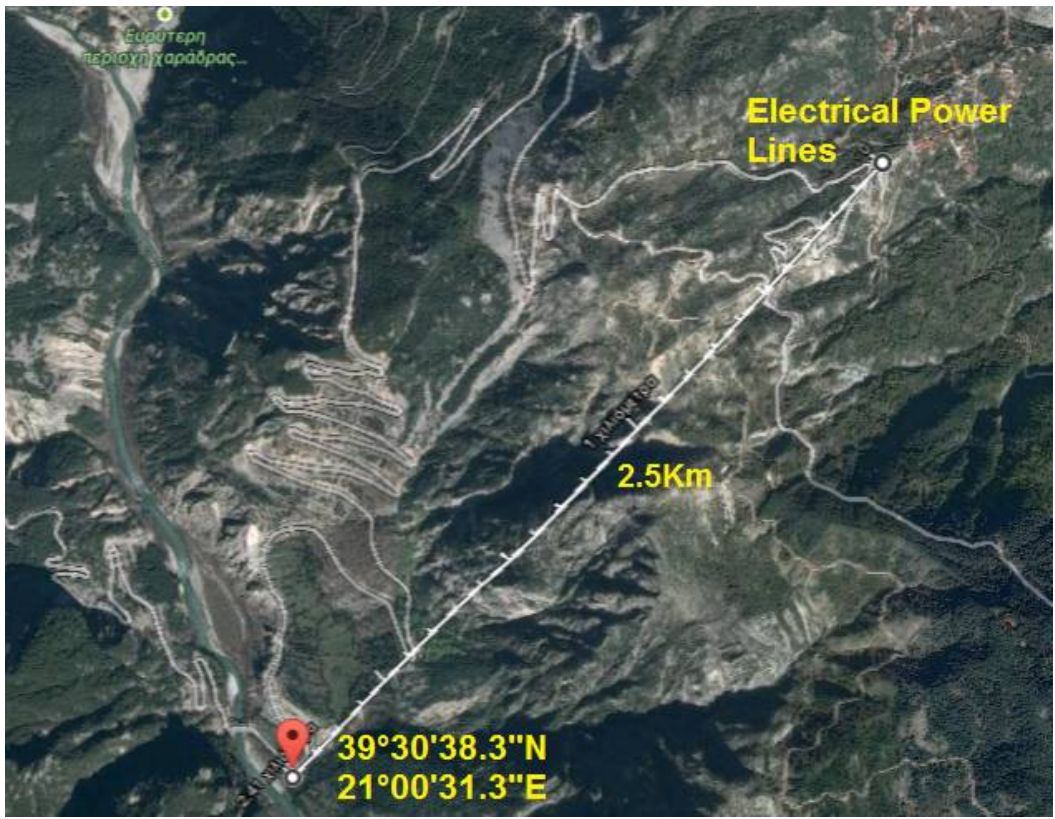

**Fig. 7. The location of the SR measurements.**

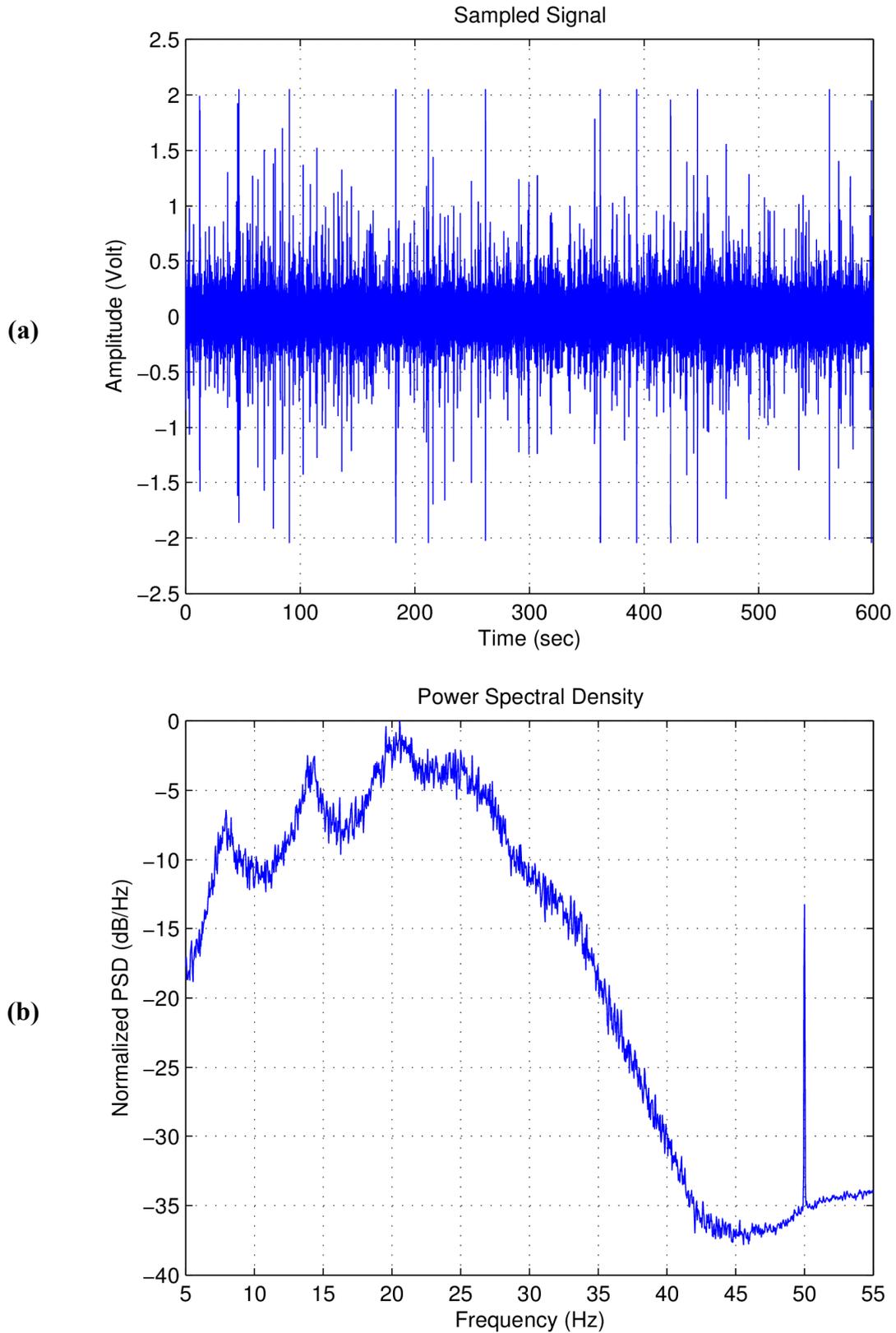

**Fig. 8. Measured (sampled) signal against time (a), and the corresponding PSD (b) using the Welch method.**

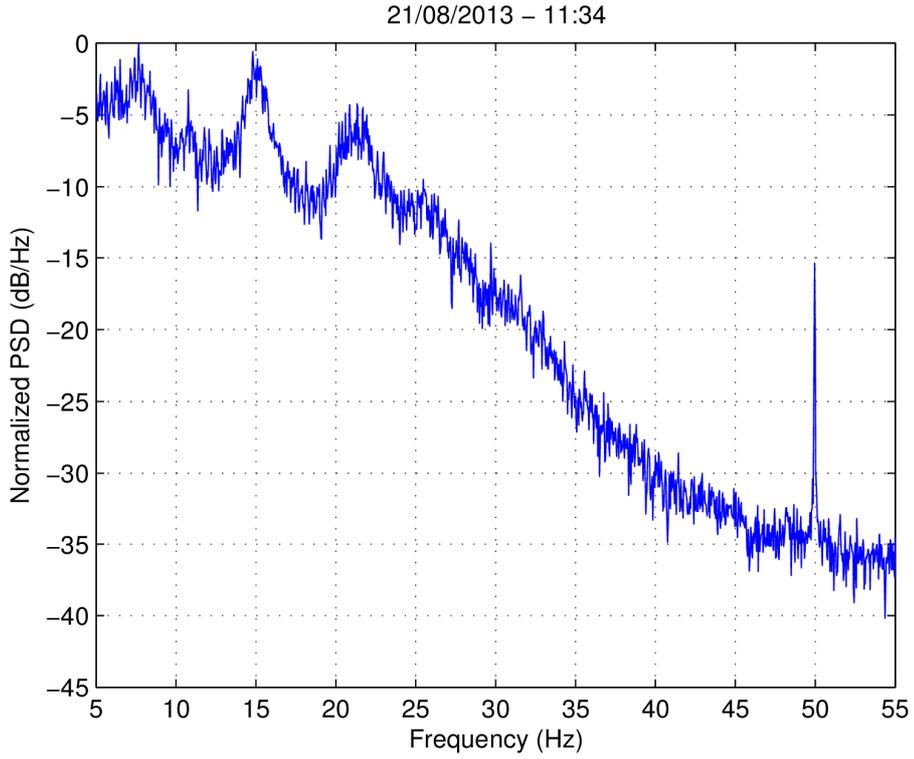

(a)

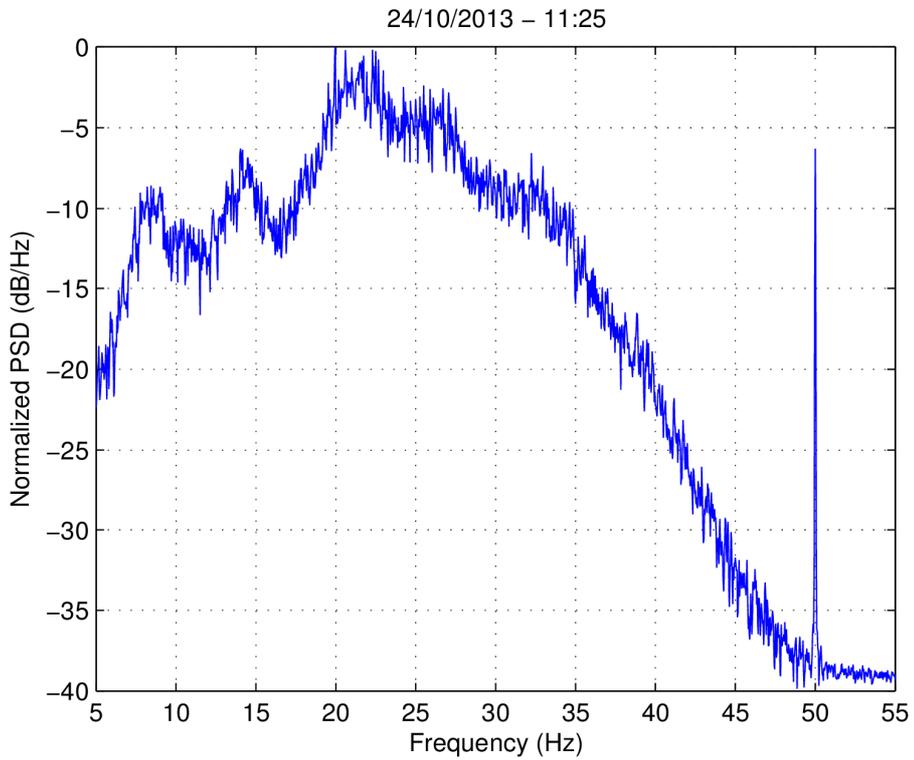

(b)

**Fig. 9. Calculated PSD using the Welch method of measured signals at dates (a) 21/08/2013, (b) 24/10/2013.**

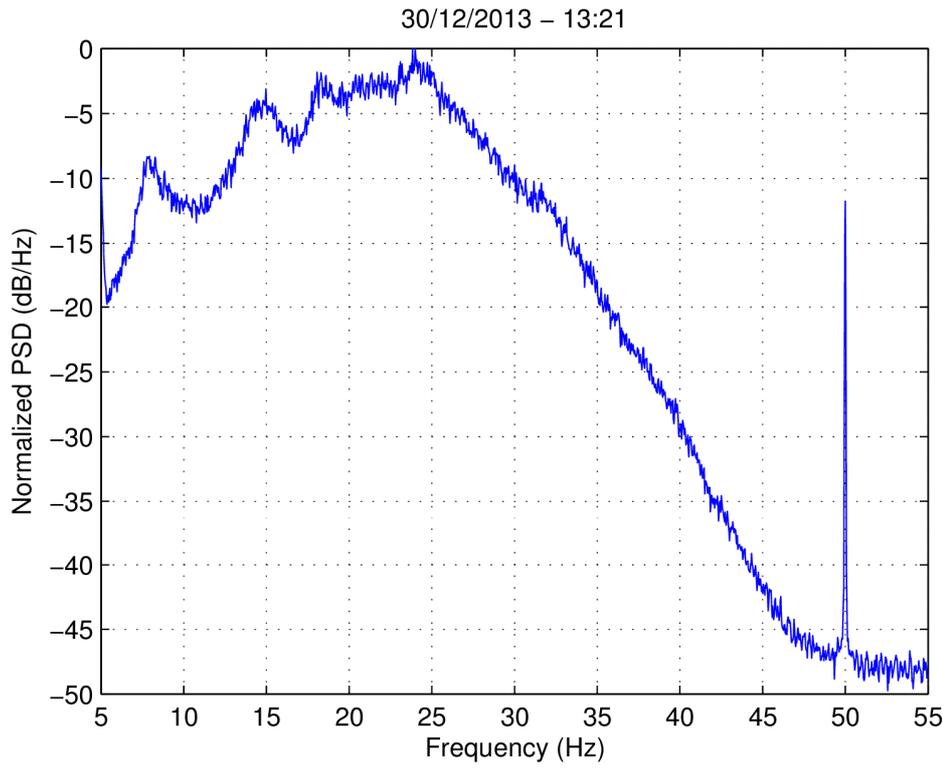

(c)

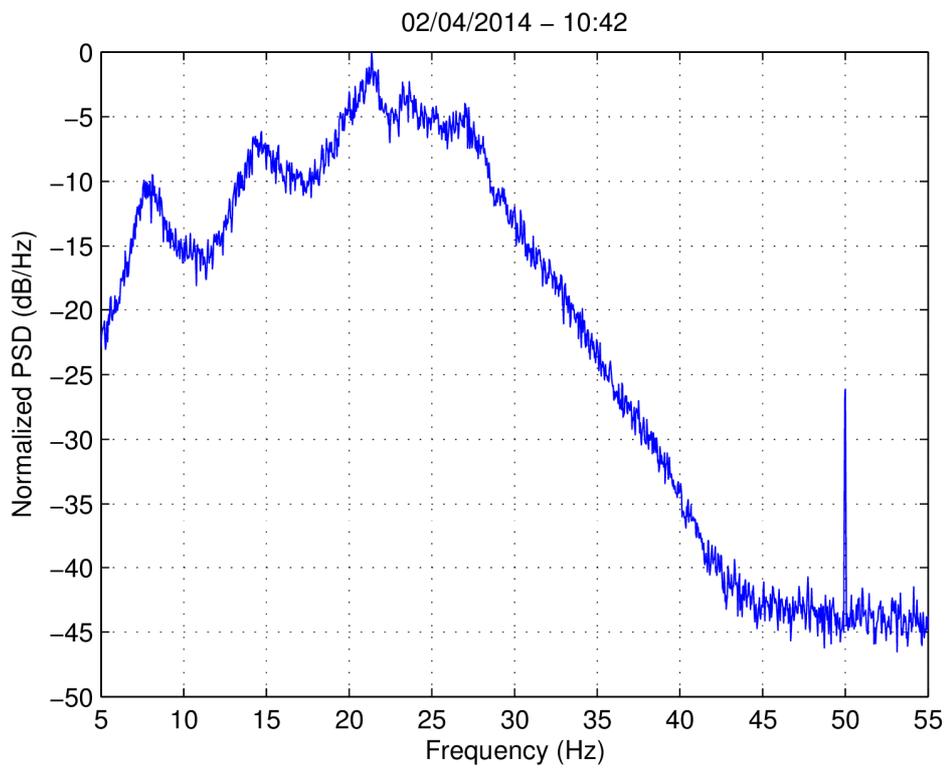

(d)

**Fig. 9. Calculated PSD using the Welch method of measured signals at dates (c) 30/12/2013, (d) 02/04/2014.**

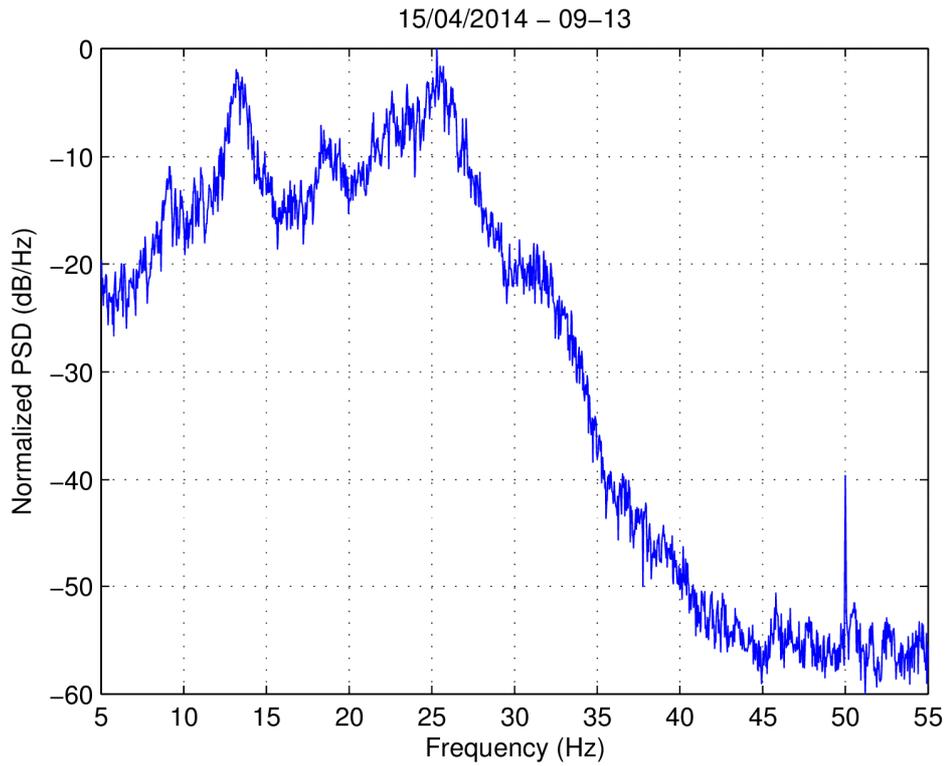

(e)

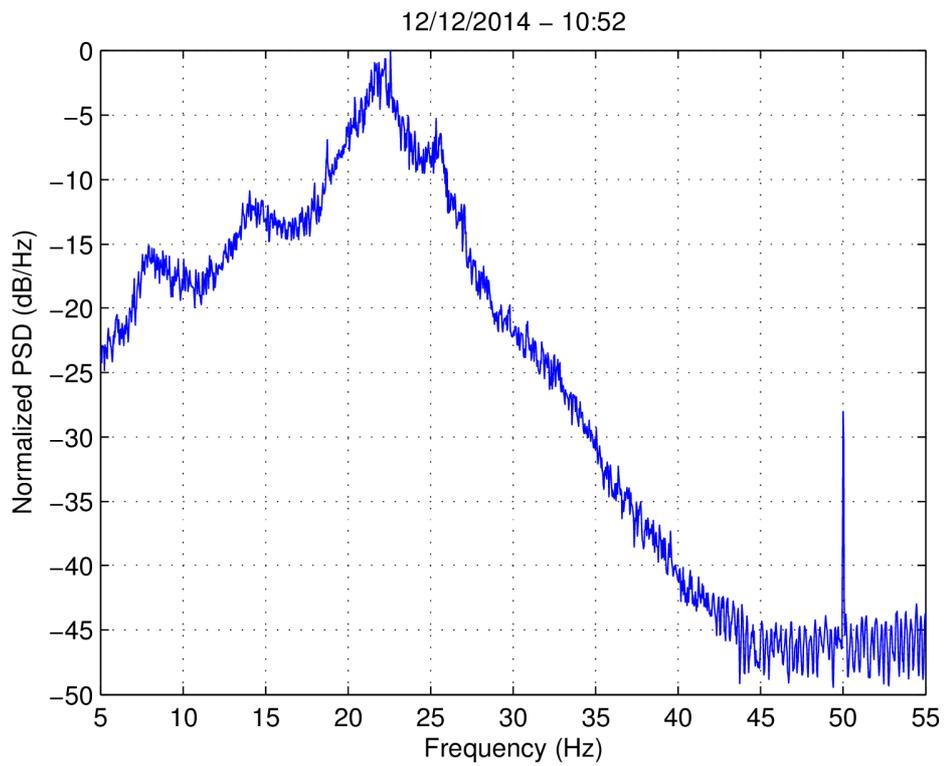

(f)

**Fig. 9. Calculated PSD using the Welch method of measured signals at dates (e) 15/04/2014, (f) 12/12/2014.**

The next step is to estimate the parameters of each resonance, which are the frequency, the magnitude and the width. For this purpose, a nonlinear least square regression, using a Lorentzian-like function for the mathematical model for the resonance estimation is performed (Sentman, 1987). The function is expressed below in Eq. 7, introducing the three aforementioned parameters,

$$P(f) = \frac{A}{1 + 4\frac{(f - f_0)^2}{W^2}} \quad (7)$$

Where, A is the peak power of the resonance, $f_0$ is the central (resonance) frequency and W represents the full width at half maximum (FWHM). Fig. 10. shows the result of the curve fitting using this model for the spectrum shown in Fig. 8. Note that for the calculations of the spectrum the power spectrum is expressed in $Watt/Hz$, whereas in Fig. 8 the units are $dB/Hz$. Obviously the whole analog front end chain containing amplifiers, various kind of filters, can not have a perfect rectangular response. Due to this an analog distortion is introduced in our signal. Knowing the front end response before any calculation, we correct this spectrum by doing equalization in the distortion by the analog front end chain. Therefore, before the calculations, equalization is used to correct the spectrum distorted by the analog front end. Also, we estimate the noise floor by performing linear least squares regression on the pure noise data. The noise level is derived by subtracting the estimated fitted resonances from the initial spectrum, under the assumption that the noise is additive and uncorrelated to the signal, in a similar technique like the well known power spectral subtraction, used in speech enhancement (Boll, 1979 ; Brouti, Schwartz and Makhoul, 1979). In order to have a measure of goodness of the results, we need to define the confidence level of our estimations. This applies to the peaks (resonances) as follows. We perform a smoothing operation on the initial noise spectrum using the moving average technique, as Fig. 11. shows. The difference between the smoothed data and the noisy spectrum is the noise variation of the spectrum. A confidence level 99% or three times the standard deviation (3*sd) is defined above the noise floor so that, each peak which exceeds this limit to be considered as confident at this level.

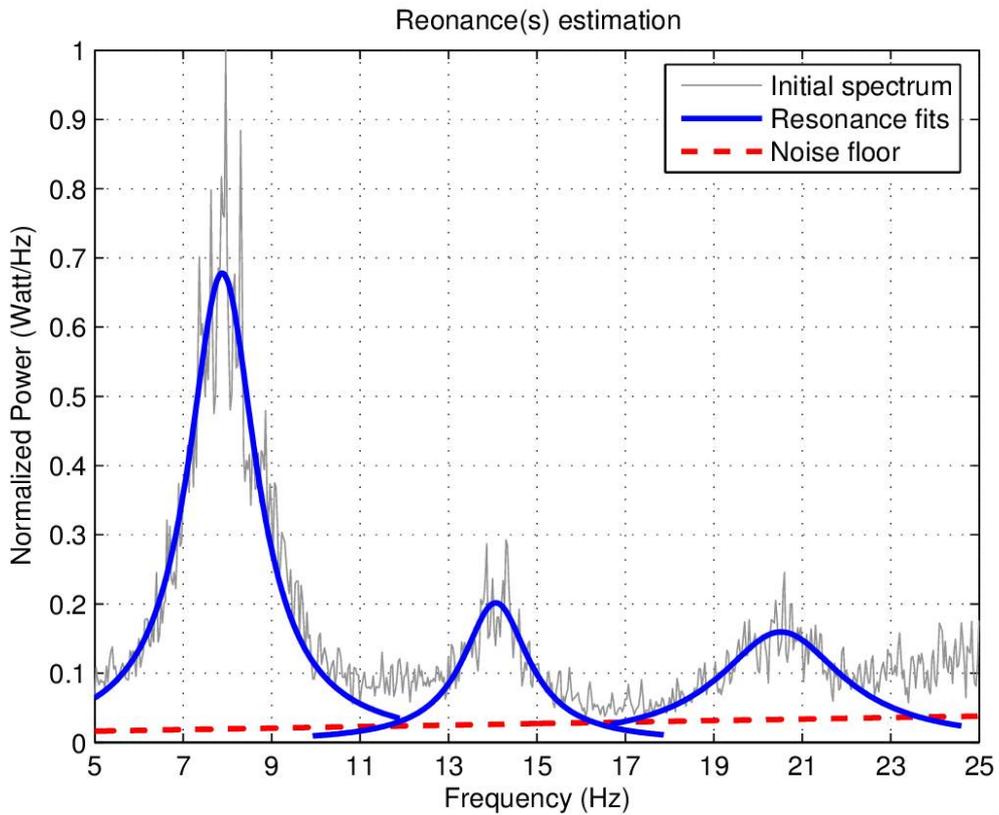

**Fig. 10. Resonance estimation by non-linear fitting using Lorentzian model.**

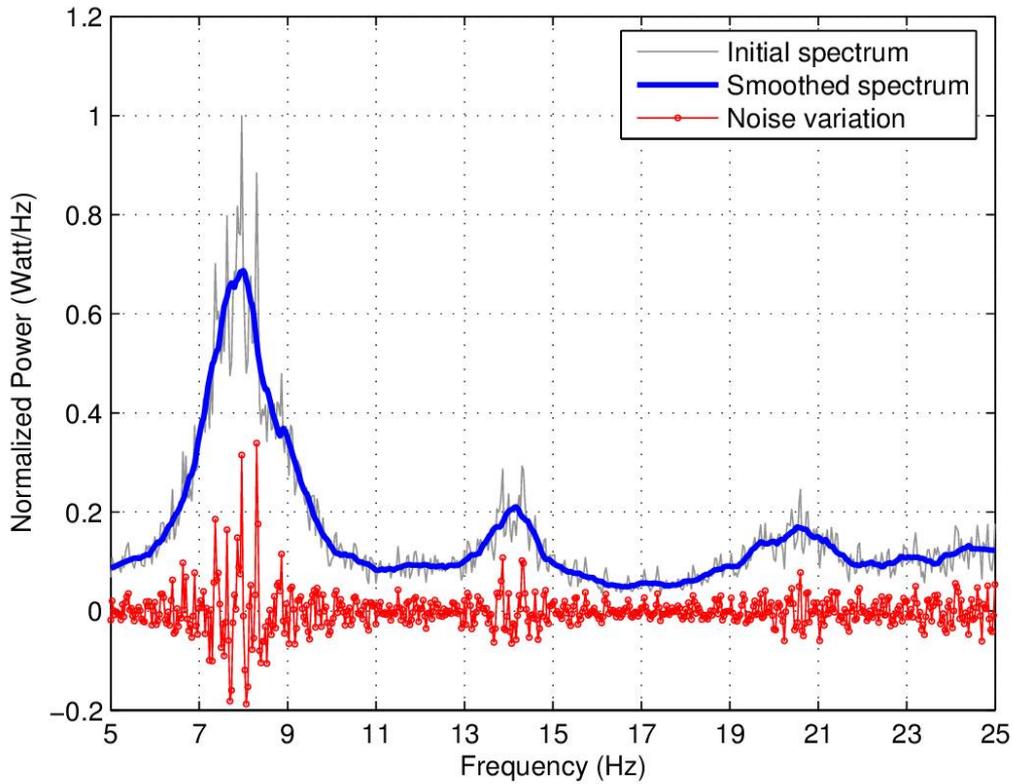

Fig. 11. Noise variation estimation to define the confidence level

6. Conclusions and future work

In this paper, a reliable, portable acquisition system for Schumann resonance measurements was presented. The system was implemented by low-cost electronics and equipment found easily in a typical research laboratory. The system uses a data logging device that records the analog signal from an antenna in association with a software application for the signal processing. The adjustable sampling rate of the data logger in conjunction with an adequate bit resolution proves out that it detects the relatively weak signal in the presence of strong noise. Accurate estimation techniques that are used to extract the spectral parameters of the Schumann resonances are analyzed in detail. Comparison between simulations and preliminary measurements confirms the correct functioning of the data acquisition and processing system. All SR measurements, within presented, for this geographical latitude and longitude are new.

In this work, we further sharing various preliminary measurements, taken occasionally from August 2013 to December 2014, coming from a prototype SR measurement system among the scientific community encouraging either individual researchers or research teams to disseminate SR measurements and results. One more SR measurement system, identical to the one presented before, has been installed and monitoring for the past month on Penteli Mountain, located east of Athens at 38°04'43.0"N, 23°56'02.5"E. We are expecting that long-term monitoring and processing of such data especially in an extremely seismic area as that of the Hellenic arc will give significant results for short-term earthquake forecasting.

Acknowledgement

We would like to thank the "Mariolopoulos-Kanaginis Foundation for the Environmental Sciences" for funding this research.

Nickolaenko, A. P., & Rabinowicz, L. M. (1995). Study of the annual changes of global lightning distribution and frequency variations of the first Schumann resonance mode. *Journal of Atmospheric and Terrestrial Physics*. doi:10.1016/0021-9169(94)00114-4

Ondrášková, A., Ševčík, S., & Kostecký, P. (2009). A significant decrease of the fundamental Schumann resonance frequency during the solar cycle minimum of 2008-9 as observed at Modra Observatory. Contributions To Geophysics And Geodesy, 39(4). doi:10.2478/v10126-009-0013-5

Palmer, S. J., Rycroft, M. J., & Cermack, M. (2006). Solar and geomagnetic activity, extremely low frequency magnetic and electric fields and human health at the Earth's surface. *Surveys in Geophysics*. doi:10.1007/s10712-006-9010-7

Proakis, J. G., & Manolakis, D. G. (1996). *Digital signal processing: Principles, algorithms, and applications*. Upper Saddle River, NJ: Prentice Hall.

Roldugin, V. C., Maltsev, Y. P., Petrova, G. A., & Vasiljev, A. N. (2001). Decrease of the first Schumann resonance frequency during solar proton events. *Journal of Geophysical Research*. doi:10.1029/2000JA900118

Schumann, W. O. (1952). Über die strahlungslosen Eigenschwingungen einer leitenden Kugel die von einer Luftschicht und einer Ionosphärenhülle umgeben ist. *Zeitschrift Fur Naturforschung Section A-a Journal of Physical Sciences*.

Sekiguchi, M., Hayakawa, M., Nickolaenko, A. P., & Hobara, Y. (2006). Evidence on a link between the intensity of Schumann resonance and global surface temperature. *Annales Geophysicae*. doi:10.5194/angeo-24-1809-2006

Sentman, D. D. (1987). Magnetic elliptical polarization of Schumann resonances. *Radio Science*. doi:10.1029/RS022i004p00595
27